# QM/MM-MD analysis of dissociation of $Ag^+$ and $H^+$ mediated cytidines.

Pavel N. Kliuev, Zakhar V. Reveguk, Petr A. Sokolov, Tomash S. Sych and Ruslan R. Ramazanov.*

St.Petersburg State University, 7/9 Universitetskaya nab., St. Petersburg, 199034 Russia.

**ABSTRACT**

We report a biased QM/MM molecular dynamics study of the dissociation process of cytidine-cytidine complexes mediated by $Ag^+$ or $H^+$ ions. We performed calculations under real solvent conditions and obtained the free energy profiles (FEP) by thermodynamic integration technique to give deep insights on the dissociation process. For all geometries corresponding to key points on FEP the noncovalent interaction descriptors (NCI) were calculated and the details of dissociation mechanism were revealed. Our findings by means of energy barrier analysis of FEP for $Ag^+$-mediated cytidines suggested more favorable cisoid over transoid configuration, in contrast to $H^+$ mediated cytidines. The existence of two energy minima in FEP, local and global, in all $Ag^+$ and $H^+$ complexes before the dissociation transition state was revealed for the first time. We showed that the global minimum of dissociation profiles for both $Ag^+$-mediated cytidine isomers does not correspond to the previously obtained QM equilibrium geometries. Our results induce future studies of longer silver mediated DNA strands comprising cytosines by QM/MM-MD.

**Introduction.**

The complexations between silver ions and nucleic acids base pairs as alternatives to H-bonding have attracted considerable interest for their involvement in various higher order DNA self-assembling nanostructures.[1–4] The common strategy is to replace the hydrogen atoms responsible for the H-bonds between nitrogenous bases of DNA by $Ag^+$ ions. This leads to a better stabilization of the DNA secondary structure due to stronger coordination bonds of nitrogenous bases with electron donor atoms. Without adding silver cations depending on solution conditions, such as pH and ionic strength, the different DNA structures can arise with canonical or non-canonical H-bonding arrangements. i-Motif DNA is a type of higher order DNA structure formed by H-bonding of cytosine-rich sequences at acidic pH. i-Motif, as suggested, can be used in many different DNA-based nanomachines.[5,6] It is assumed that, in contrast to cytosine-cytosine H-bonds promoted by acidic conditions, the same DNA structure can be folded using $Ag^+$ ions even at neutral pH possessing much stronger bonding by C-$Ag^+$-C metal-mediated base pairing.[7] To date there are numerous experimental studies devoted to construction of the DNA diverse forms by using C-Ag[+]-C mismatch, in which Ag[+] ions provide DNA assemblies with enhanced stability.[1,2,7–19] It was shown that DNA can be folded with Ag[+] by at least two ways, forming parallel[20,21] or anti-parallel[10,22] duplex strand orientation. Two distinct geometries of the opposing cytidines, corresponding to formation of parallel or anti-parallel DNA duplex, were denoted as transoid (H[+] or Ag[+]) and cisoid (H[+] or Ag[+]), respectively, regarding to rotation around N3-N3 direction (Fig. 1).

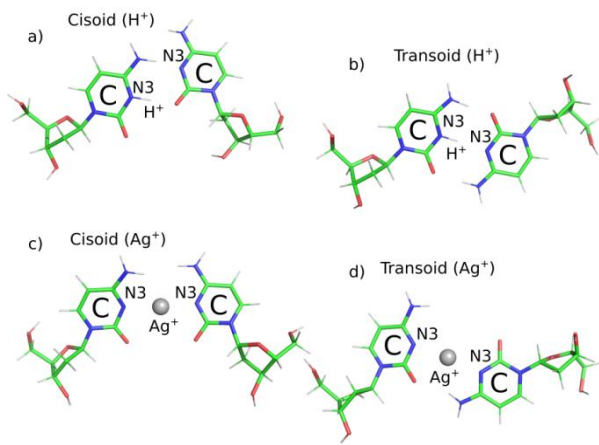

Figure 1. Equilibrium geometries of studied cytidine complexes mediated by Ag[+] or H[+]: a) cisoid (H[+]), b) transoid (H[+]), c) cisoid (Ag[+]), d) transoid (Ag[+]).

Both the transoid and cisoid C-Ag[+]-C pairing structures were observed in X-ray crystallography and NMR experiments,[9,18,19,23,24] but to date any reliable arguments, why one configuration is preferable to another, are not presented. Theoretical considerations using dispersion-corrected DFT calculations of interaction energies at stationary points of isolated C-Ag[+]-C pairs in water solvent revealed that both rotational isomers are approximately isoenergetic.[4] In the case of parallel double-helix silver-mediated cytosine homobase strands, the interplane H-bonds were suggested as an additional stabilization factor.[14,17] However, at present, the molecular level explanation of the possible dynamic pathway for C-Ag[+]-C pair formation is not presented. The previous QM equilibrium geometries[4] of isolated silver-mediated cytidine pairs are very close to those found in real crystallography or NMR structures.[9,18,19,23,24] Therefore, these geometries are considered relevant to the experiment. But such a generalization deserves a more detailed analysis because the environment in which the complex is located can have a noticeable effect on it. Also the present study is aimed to solve the following issue: are there fundamental differences in the dynamic of cisoid or transoid cytidine-cytidine (C-C) complex formation by Ag[+] or H[+] ions? We performed detailed comparative theoretical analysis for dissociation process kinetics of isolated C-C pairs in cis- and trans- orientations mediated by $Ag^+$ or $H^+$ ions under real solvent conditions using a multiscale technique combining quantum and molecular mechanical representations (QM/MM). For the silver-mediated cytidines, our findings by means of energy barrier analysis suggest more favorable cisoid over transoid configuration in the dissociation process. We have shown for the first time that the QM/MM local but not global minimum of dissociation profiles for both isomers of single C-C pairs mediated by $Ag^+$ corresponds to the geometries found in the experiment where extended DNA was considered. This hints that the presence of additional stabilization factors in extended metal-mediated DNA that are not taken into account in our model of an isolated nucleotide pairs shifts the system from the global minimum to the local one. That states the real picture of DNA self-assembling by silver ions is more complicated and for the deep understanding of studied systems is crucial to account for extended polymer structure and metallophilic interaction of neighbor silver ions. In comparison to $Ag^+$, the C-C pairs mediated by $H^+$ ion have an opposite situation and, as expected, possess more short-range character of interaction and smaller energy barrier in the dissociation process.

**Theoretical methods.**

The biased QM/MM molecular dynamics simulations were performed and free energy profiles (FEP) were obtained by thermodynamic integration technique to discover the details of dissociation process of two cytosines mediated by $Ag^+$ or $H^+$ ions. In order to accomplish this, at the first stage, rough classical molecular dynamics, pulled in direction associated with plane containing N3-($Ag^+$ or $H^+$)-N3 bonds, was made for achieving the dissociation between cytidines. In our model, we used cytidine instead of cytosine to account for the effect of sugar movement, as it happens in real DNA in a rough approximation. It was necessary to obtain a well equilibrated solvent around the reagents as a good starting point for subsequent QM/MM simulations. Next, we cut out the frames of classical MD trajectories at each step of 0.1 A from the initial configuration (the distance is counted between N3 cytosine atoms). For each frame, constrained QM/MM geometry optimization with frozen N3-N3 distances was performed to allow the quantum chemical rearrangements of simplified classical models. Then, starting from these configurations, the biased QM/MM molecular dynamics simulations using constraints for N3-N3 distances by the Shake algorithm[25] were performed within canonical (NVT) ensemble at 300 K with 0.5 fs time step. Employing these simulations, the average Lagrange multipliers of the Shake algorithm were calculated and the dissociation FEPs were constructed. For a more detailed study, the calculations of noncovalent interaction descriptors (NCI) were carried out to characterize and generalize the patterns of binding process.

All classical calculations were carried out in Gromacs v. 5.0.6 package[26] using the latest Amber family force field parmBSC1[27] adapted for silver ions and DNA complexes (see SI). We performed classical MD, pulling over the course of 500 ps of one cytidine from another by breaking a single $Ag^+$-N3 bond and applying harmonic potential between centers of mass. For this purpose, the force constant of 100 kJ mol$^{-1}$ nm$^{-2}$ and the rate of 0.01 nm per ps, at which the imaginary spring attached to our pull groups is elongated, were used. The MM model scene included a solvent (water TIP3P) cubic box 4x4x4 nm and periodic boundary conditions. All trajectories were obtained in NVT ensemble using the Parrinello-Raman thermostat at T=300 K and time step of 1 fs. All QM/MM calculations were performed in CP2K v 2.6 package.[28] The QM/MM interface used an additive Hamiltonian[29] and comprised QM and MM regions connected by H-link atoms. The noncovalent interactions between QM and MM parts were represented by electrostatic interaction using Gaussian expansion of the electrostatic potential method (GEEP)[29] employing the electrostatic coupling procedure. QM region, including the $Ag^+$ or $H^+$ mediated nitrogenous bases of cytosines, was treated at the DFT level, whereas the comprising solvent molecules and sugars MM part was described by modified parmBSC1 force field. In the case of $Ag^+$ mediated cytidines, the QM part was 12.34x20.25x8.08 Å$^3$, while the QM part of the system containing $H^+$ ion mediated cytidines was 11.44x17.43x8.95 Å$^3$. The QM box without periodic boundary conditions was set to the size corresponding to the area covering all QM-atoms during the whole QM/MM-MD trajectory. The DFT calculations used the combination of the Gaussian and plane-wave (GPW)[30,31] scheme for electronic density representation, double-ζ valence plus polarization (DZVP) basis sets of the MOLOPT[32] type to describe the valence electrons and norm-conserving Goedecker-Teter-Hutter (GTH)[33–35] pseudopotentials to approximate the core electrons, the exchange and correlation Perdew−Burke−Ernzerhof (PBE)[36] functional with Grimme's D3 dispersive correction[37] to account for long-range van der Waals interactions in QM part. The same QM/MM arrangement was successfully used for similar systems.[17, 38–40] NCI descriptors were calculated using the NCIPLOT program.[41]

**Results.**

An important point in the structure geometry calculation and dissociation energy profile in metal-organic frameworks is the choice of an appropriate quantum-chemical (QM) approach allowing cost-effective calculations with acceptable accuracy.[42] In this case, the VdW-corrected

PBE functional as DFT exchange-correlation approach is the most advantageous choice since it can be expanded to study DNA-metal structures with good accuracy. In our QM/MM models containing $H^+$ and $Ag^+$, a previously adapted PBE-D3/parmBSC1 scheme was used to study geometry of luminescent partially oxidized silver clusters stabilized on DNA.[40] Our biased QM/MM-MD gave unusual results lying in a slight inconsistency with the previous pure MD and QM studies. The existence of the two minima in energy, local and global, was revealed in all trajectories before the transition state corresponding to dissociation of the complexes. In Fig. 2, the smoothed curves display the PMF (kcal/mol) variation as a function of N3-N3 separation distance (Å), reflecting the main differences and similarities in the processes.

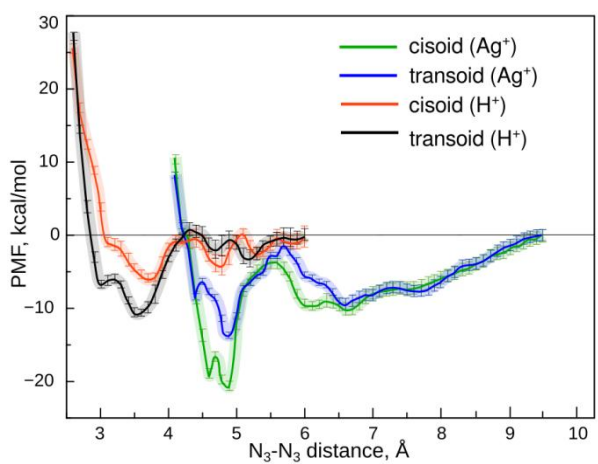

Figure 2. Smoothed curves of the dissociation free energy profiles of $Ag^+$ and $H^+$ mediated cytidines with statistical errors indicated by the error bars.

In the case of $H^+$ ion mediated cytosines, the transoid has two first energy minima at 3.0 Å and 3.5 Å, and the cisoid has also two minima at 3.1 Å and 3.6 Å. The first energy minimum of cisoid ($H^+$) is not so deep, but the energy of 5.97 kcal/mol was obtained for the second minimum, which is comparable to different mismatch base-pairings containing two H-bonds previously studied by Vendeix et al.[43] For transoid ($H^+$), unexpected shifts of the first local minimum were observed from the distance of 3 Å, corresponding to a complex with three standard H-bonds, towards a more favorable configuration at 3.5 Å with increasing in energy ca. 4 kcal/mol. A careful look at the complex structure revealed the base pair stabilization by the water bridge at 3.5 Å (Fig. 3), which looks like an artifact that strongly affects the FEP shape. Even so, the existence of low and slightly sloped energy barrier between two minima can be attributed to steric hindrance of H-bonds within the QM/MM-MD trajectory, which is more precisely described in QM representations than in pure MM. Silver-mediated cytosines also have two energy minima with different distances. The transoid ($H^+$) has two minima at 4.4 Å and 4.9

Å, while the cisoid (Ag$^+$) has two minima at 4.5 Å and 4.9 Å. It should be noted that many authors consider the equilibrium configurations with the distance ca. 4.4 Å as the lowest energy geometry close to revealed in crystallography and NMR in the DNA double helix.[17,42–45] Here we have shown that more favorable configurations exist for single metal-mediated cytidine pairs in water solvent.

*H$^+$ mediated cytidines profile.*

As was mentioned above the dissociation profiles have an extraordinary character for both rotational configurations of H$^+$ mediated cytidines, but there are some differences between them. For a subtler assessment of the binding characteristics, NCI descriptors for all key geometries were calculated using electron density.

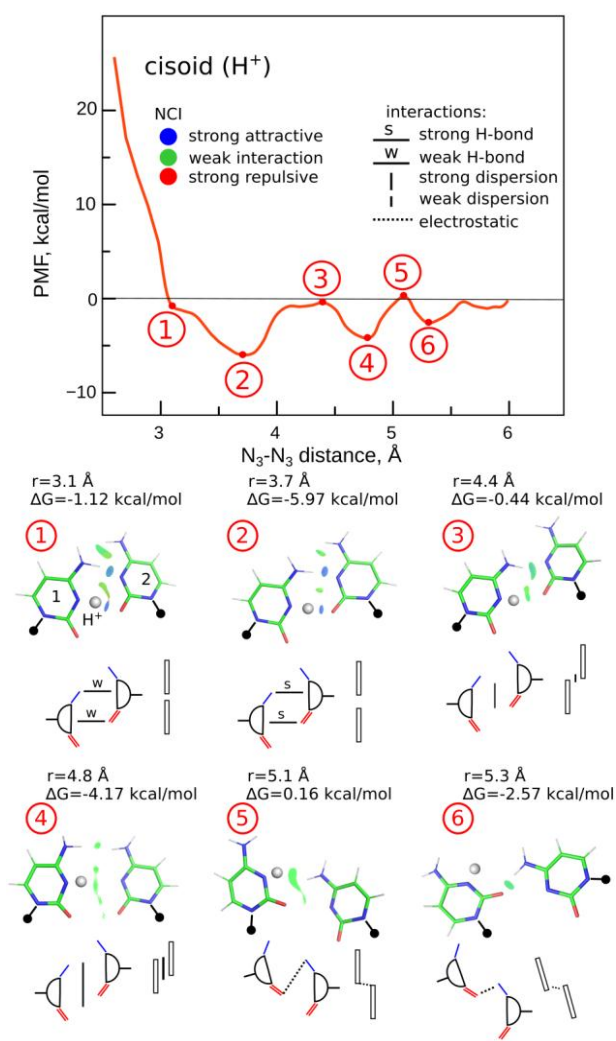

Figure 3. Detailed illustration of the dissociation profile of cisoid (H$^+$) with depicted character of bonding. For clarity sugars are omitted. NCI descriptors are shown.

The point **1** at 3.1 Å and point **2** at 3.6 Å of cisoid (H$^+$) FEP shown in Fig. 3 correspond to the local and global minima that were described above. The structures **1** and **2** look very similar and include two H-bonds N-H...N and N-H$^+$...O. For **1** the distance $r$(N-H...N) is 2.82 Å and distance $r$(N-H$^+$...O) is 2.69 Å, while for **2** the distance $r$(N-H...N) is 2.80 Å and distance $r$(N-H$^+$...O) is 2.77 Å. However, **2** is more favorable than **1** up to 4.65 kcal/mol. The reason is that a steric hindrance of H-bonds and electrostatic repulsion of the exocyclic amino groups in **1** is more significant and leads to a destabilization, as shown by NCI in Fig. 3. Therefore, in this case, we denoted H-bonds in **1** as "weak H-bonds" and H-bonds in **2** as "strong H-bonds". Next, the point **3** corresponds to the transition state (TS), which is formed by displacement of the base planes into the partial stacking geometry. This state is denoted as a "weak dispersion" interaction, and the point **4** is the more favorable state that looks like geometry with "strong dispersion" interaction. Further increase in the distance between the cytosines leads to rotation of one of cytosines around the centers of mass so that the transition state **5** is formed with a "weak electrostatic" interaction between the exocyclic carbonyl and amino groups. Then a more favorable state **6** with a "strong electrostatic" interaction is formed. For transoid (H$^+$) configuration, **1** and **2** FEP points also had the energy difference due to water bridge, as shown in Fig. 4. Stabilization by water molecule gave additional two H-bonds and reduced free energy to 10.57 kcal/mol, which is twice the standard value of the free energy of canonical and non-canonical base pairs. Schematic representation of bonding rearrangements is illustrated in Fig. 4. Further transformations of the complex occur in the same manner as in the cisoid (H$^+$) case for points **3**, **4**, **5**, **6**. In general, transoid (H$^+$) orientation is clearly preferable to the cisoid (H$^+$), and this was expected.

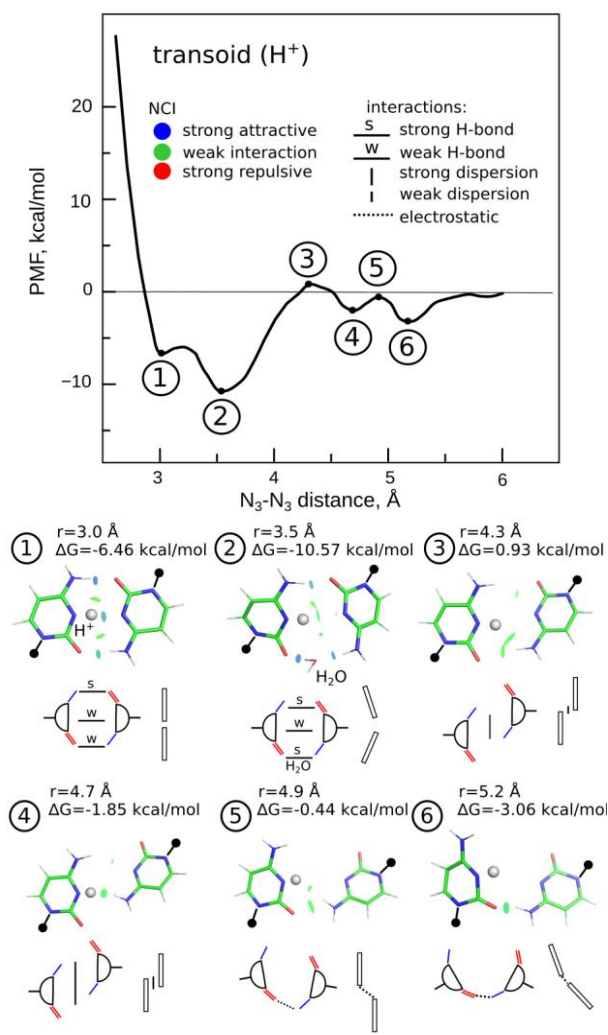

Figure 4. Detailed illustration of the dissociation profile of transoid (H$^+$) with depicted character of bonding. For clarity sugars are omitted. NCI descriptors are shown.

*Ag$^+$ mediated cytidines.*

Dissociation profiles for Ag$^+$ mediated cytidines have a similar trend for both isomers. The first local energy minimum is the point **1** of the cisoid (Ag$^+$) located at 4.5 Å (Fig. 5) and the point **1** for the transoid (Ag$^+$) located at 4.4 Å (Fig. 6). It should be noted that the cisoid (Ag$^+$) has the same energy as the transoid (Ag$^+$) at 4.4 Å, but there are no barriers for cisoid up to 4.5 Å. Many theoretical studies of equilibrium geometries of trans- and cis- configurations of silver-mediated cytosines determine the equilibrium values of N3-N3 distance in the range of ca. 4.20-4.50 Å depending of the method and solvent used.[12,42,44,46,47] NMR and X-ray crystallography vary this value in the range of ca. 4.24-4.72 Å (2RVP.pdb:[24] cisoid *r*=4.42 Å; 5AY2.pdb:[48] cisoid *r*=4.35 Å, 4.55 Å; 5IX7.pdb:[13] cisoid *r*=4.72 Å, 4.36 Å; 5XJZ.pdb:[16] transoid *r*=4.25 Å; 6NIZ.pdb:[23] transoid *r*=4.24 Å).

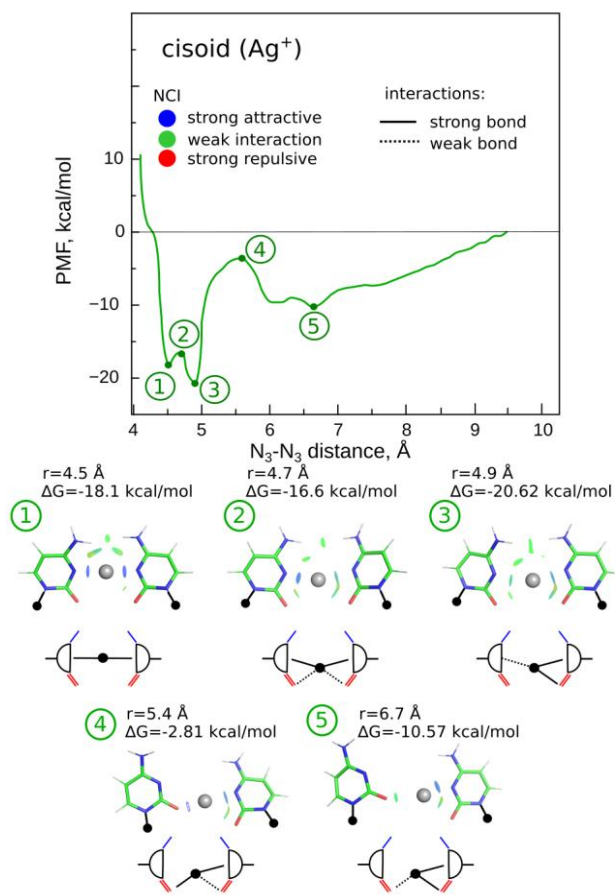

Figure 5. Detailed illustration of the dissociation profile of cisoid (Ag$^+$) with depicted character of bonding. For clarity sugars are omitted. NCI descriptors are shown.

Obviously, the local geometry of N-Ag-N strictly depends on an outer polymer or solvent, which affect the coordination behavior of silver in relation to lone pairs of nitrogens or oxygens. Thus, in this sense, the following assumption seems reasonable: DNA self-assembling by silver ions includes various factors, among which the cytosines pairing by metal ions is one of the main drivers but not the only factor for the supramolecular structure stabilization. The point **2** of the FEP for both isomers relates to a minor barrier not exceeding 3 kcal/mol until they don't fall into the global energy minimum In the case of the cisoid (Ag$^+$), the TS corresponds to four weak coordination bonds with exocyclic carbonyl and amino groups of both cytosines holding a flat geometry, while the transoid (Ag$^+$) is characterized by a slight rotation along the axis of N3 bonds. The next point **3** corresponds to the global minimum with energy difference ca. 7 kcal/mol for isomers, wherein the cisoid (Ag$^+$) lies lower than the transoid (Ag$^+$) on the FEP. This happens due to a slight rotation in transoid (Ag$^+$) leading to appearance of three "weak bonds" versus two "strong bonds" and one "weak bond" in the cisoid (Ag$^+$). The following two points **4** and **5** for both isomers suggest the continued dissociation by creating strong and weak electrostatic interactions through rotation of one of cytosines as in the case of H$^+$ mediated nucleotides described above.

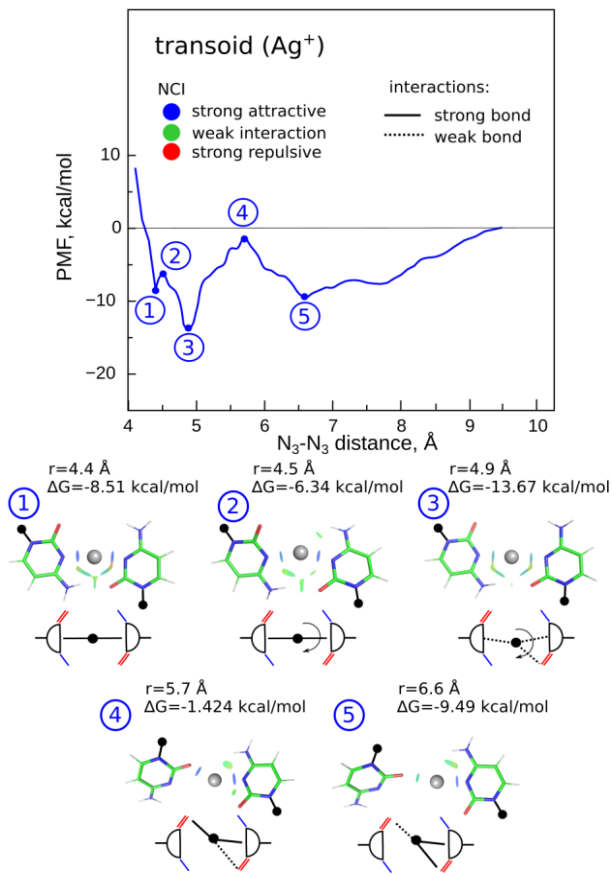

Figure 6. Detailed illustration of the dissociation profile of transoid (Ag$^+$) with depicted character of bonding. For clarity sugars are omitted. NCI descriptors are shown.

In general, it should be borne in mind that the obtained profiles may differ regarding on the number of adjacent metal-mediated base pairs or exact polymer sequence supporting self-assembling by different factors like H-bonds or stacking.

**Conclusions.**

To the best of our knowledge, we performed FEP analysis of dissociation process of silver-mediated cytosines by applying QM/MM technique for the first time. We have shown that the silver-mediated cytosines dissociate differently from those stabilized by H$^+$. This feature should certainly be taken into account when silver ions are considered as alternative and an improving agent for the folding of the DNA nanostructures. Evidently, the behavior of base pairs in the vicinity of the global minimum leaves many questions about the contribution of various factors leading to DNA self-assembling. It becomes obvious that in order to understand the complexity of such chemical transformations, it is not enough to consider only the energy difference between the equilibrium geometries at QM level as was suggested earlier. To the same extent, it is not very informative to consider large DNA-$Ag^+$ systems purely in the classical molecular mechanical representation, since they do not take into account possible metallophilic interactions. The factors such as stacking interactions between neighboring nucleobases, interplane H-bonds as well as metallophilic interactions between adjacent metal-mediated base pairs can have a complex impact on the overall stability. The results obtained in the present study prompt further efforts in studying more complex DNA-$Ag^+$ systems containing several silver mediated base pairs using developed QM/MM-MD approach. We believe that a more in-depth study of the interactions of silver ions with nucleotides by proposed multiscale technique will rationalize experimental approaches in the selection of strategies for folding DNA nanostructures.

ACKNOWLEDGMENT

This work was supported by the Russian Science Foundation (project 17-73-10070). The reported calculations were performed at the Supercomputing Center of Lomonosov Moscow State University and at the Supercomputing Center of St. Petersburg State University.